# Platinum Disulfide (PtS$_2$) and Silicon Pyramids: Efficient 2D/3D Heterojunctions Tunneling and Breakdown Diodes

Sikandar Aftab[1], Ms. Samiya[2], Muhammad Waqas Iqbal[3], Fahmid Kabir[1], Muhammad Zahir Iqbal[4], M. Arslan Shehzad[5*]

[1]Department of Engineering Science, Simon Fraser University, Burnaby, BC, Canada

[2]Department of Environment and Energy, Sejong University, 209 Neungdong-ro, Gwangjin-gu, South Korea

[3]Department of Physics, Riphah International University, 14 Ali Road, Lahore, Pakistan

[4]Nanotechnology Research Laboratory, Faculty of Engineering Sciences, GIK Institute of Engineering Sciences and Technology, Topi 23640, Khyber Pakhtunkhwa, Pakistan

[5]KECK II Facility, NUANCE Center and Department of Material Science and Engineering, Northwestern University, Evanston, Illinois 60208, United States

Corresponding E-mail: arslan@northwestern.edu

**Abstract:** The p-n junction constructed from the group-10 TMDCs, or namely, transition metal dichalcogenides with an intrinsic layered structure, is not considerably reported. This study presents a mechanical exfoliation-based technique to prepare $PtS_2$/pyramids Si *p-n* junctions for an investigation of the tunneling and breakdown diodes. The demonstrated *p-n* diode exhibited a high rectifying performance reaching a rectification ratio ($I_f/I_r$) to ~7.2 ×$10^4$ at zero gate bias with an ideality factor of ~1.5. The Zener tunneling was observed at a low reverse bias region of breakdown voltage (from -6 to -1.0 V) at various temperatures (50 to 300 K) and it was a negative coefficient of temperature. Conversely, for the greater breakdown–voltage regime (-15 to -11 V), the breakdown voltage increased with the increased temperature (150 to 300 K), indicating a positive coefficient of temperature. Therefore, this phenomenon was attributed to the avalanche breakdown. The *p-n* junctions displayed photovoltaic characteristics under the illumination of visible light (500 nm), such as high responsivity ($R_{ph}$) and photo gain (G) of 11.88 A/W, and 67.10, respectively. The maximum values for both the open-circuit voltage ($V_{OC}$) and the short-circuit current ($I_{SC}$) were observed to be 4.5 V, and 10 µA, respectively, at an input intensity of light 70.32 mW/$cm^2$. The outcomes of this study suggest $PtS_2$/pyramids Si *p-n* junctions may be employed in numerous optoelectronics including photovoltaic cells, Zener tunneling diodes, avalanche breakdown diodes, and photodetectors.

# 1.0 Introduction:

The transition metal dichalcogenides have shown many promising applications when it comes to the next generation applications, including switching photodiodes,[1, 2] photodetectors[1, 3-5], Spintronics[6], field-effect transistors,[7] piezotronics,[8] and superconductors.[9] The bandgap of many transition metal dichalcogenides range from 1 eV to 2 eV.[7, 10] Particularly, the TMDs based on molybdenum and tungsten are promising materials due to their prominent optoelectronics and semiconducting properties with ultrathin nature.[11, 12] Other 2D materials such as platinum disulfide ($PtS_2$), $PdSe_2$ and $PtSe_2$ are emerging as the most exciting TMD material due to their high mobility features.[12-17] $PtS_2$ can be transformed into monolayers or multilayers or in bulk form through a technique named mechanical exfoliation, which is done using an adhesive tape and shows a bandgap of 0.3 eV in the bulk form.[14] The structure of the $PtS_2$ monolayer is formed by the in-plane S-Pt-S atoms in which two layers/sheets of S atoms sandwich one layer/sheet of Pt atoms packed in a hexagonal structure.[18] In the bulk form of $PtS_2$ different layers are stacked together under the action of weak intermolecular forces which produce weak attraction between different layers.[18] $PtS_2$ is in focus due to its potential applications in optoelectronics and photonics for future generations. $PtS_2$ is becoming a novel participant in comparison to conventional semiconductors and graphene with remarkable features for optoelectronics, chemical sensors, field-effect transistors, and memory devices, etc. The p-n junctions lay foundations of many promising semiconductor devices in the optoelectronics industry including photovoltaic devices, photodetectors, bipolar transistors, and LEDs. Several reports support that 2D materials can be grown or exfoliated using $WS_2/Si$,[19] $MoS_2/Si$,[20-22] $MoS_2/GaTe$,[23] and $MoS_2/lnP$,[24] etc. However, the p-n junctions based on a typical group-10 TMDs with an intrinsic layered structure with Si are not considerably reported in the literature. Maintaining consistency in TMDs during their growth is notoriously difficult, and therefore, p-n junctions based on it do not show good performance in both photovoltaic and electrical characteristics.

In this study, we revealed high-performance $PtS_2$/pyramids Si heterojunctions based on mechanical exfoliation. The p-n junctions were developed by the intercalation of the multi-layer $PtS_2$ nanoflakes on the top of a p-type pyramids silicon monolayer when an oxide layer is not present. We have also confirmed that p-n junctions show improved electrical properties by annealing for 4 hours at 200ºC in a tube furnace under $Ar/H_2$ (97.5% Ar, 2.5% $H_2$) atmosphere. This research offers a mechanical exfoliation-based technique to make $PtS_2$/pyramids Si *p-n* junctions for an investigation of the band-to-band tunneling as well as avalanche effect in diodes. A PDMS stamp was used to exfoliate multi-layer $PtS_2$ nano-flakes on the top of p-type pyramids Si monolayer. The demonstrated *p-n* diodes exhibited an ideality factor of ~1.5 and high-performance rectification ratio ($I_f/I_r$) of about ~7.2 ×10$^4$. The Zener breakdown voltage was observed showing a negative coefficient of temperature at low reverse bias regime from -6 to -1.0 V at when temperature changes, i.e., from 50 to 300 K. On other hand, for the greater voltage regime from -15 to -11 V, breakdown–voltage increased in the temperature range from 150 to 300 K, indicating a positive coefficient of temperature for avalanche breakdown diodes. The devices displayed various photo response parameters under illumination of light with a responsivity and photo gain of 11.88 A/W, and 67.10, respectively. The maximum values of open-circuit voltage was 4.5 V, whereas for the short-circuit current it was 10 µA. This study proposes several optoelectronic applications, such as photovoltaic cells, LEDs, Zener diodes, breakdown diodes, and photodiodes.

## 2.0 Experimental section:

Multi-layer $PtS_2$ nanoflakes were employed to fabricate p-n heterojunction over a silicon (Si) substrate without oxide layer. P-type Si substrates were cleaned with methyl alcohol and acetone, followed by desiccation with nitrogen flow. The resistivity of the substrate lies between 1 to 10 $\Omega$.cm. The $PtS_2$ nanoflakes are obtained from bulk crystals and exfoliated using mechanical exfoliation method and dry transferred on the surface of Si with a PDMS stamp. $PtS_2$ flakes were examined with an optical microscope. The optical transparency of the $PtS_2$ nanoflakes was changed due to the difference in layer thickness. Furthermore, the thickness of the $PtS_2$ flakes, topography, and morphology characterizations were explored using Raman spectroscopy, AFM, and optical microscopy, respectively. The spot size of the laser for Raman spectroscopy was chosen at 0.7 µm with a wavelength of 514 nm. To restrict the induction of the laser-generated thermal effects, the laser beam power was kept less than 0.1 mW. We fabricated $PtS_2$/pyramids Si p-n junction devices by atomic layer deposition (ALD), photolithography and electron beam lithography. Larger electrodes were synthesized using photolithography to deposit the $Al_2O_3$ thick layer of around 25 nm thickness by using ALD to avoid the leakage current between electrodes and Si. Later on, the samples were placed inside a thermal evaporator to depose Cr/Au (6/30 nm) film. To obtain the top fine electrodes on $PtS_2$ nanoflakes, the electron beam lithography process was performed to pattern the electrodes to depose Au (80 nm) film from large electrodes to the top of the flake. After fabrication of the p-n junction device, all devices were annealed for 4 h at 200°C in the furnace tube underflow of $Ar/H_2$ (97.5% /2.5%) gas to get clean interfaces. Electrical transport measurements were taken at room temperature and low temperatures under a vacuum. The photoresponse performance of the device was measured using a visible light having wavelength of 500 nm with various input power using the exposure system.

## 3.0 Results and Discussion:

Before the device fabrication was initiated, the bulk crystals of $PtS_2$ were exfoliated on the top of a textured pyramids silicon substrates (In Figure S1 of supporting information) through a crystalline polydimethylsiloxane stamp, and the multilayer $PtS_2$ flake was investigated using the optical microscope. Figure 1(a) illustrated the structural configuration the devices, while Figure 1(b) displays the optical image of the $PtS_2$/pyramids Si p-n diode. Figure 1(c) displays the absorption spectrums of the $PtS_2$ and pyramids Si/ $PtS_2$ p-n diode. It can be seen the absorption increased by exfoliating $PtS_2$ nanoflakes on the surface of pyramids Si. Figure 1(d) displays the Raman spectroscopy of a $PtS_2$ nanoflake on a p-type silicon substrate. Two Raman signatures of $PtS_2$ flake were examined at 307.8 $cm^{-1}$ and 341.2 $cm^{-1}$ both corresponding to phonon modes $E^1_g$ and $A^2_{1g}$, respectively. Variation in the two modes is 33.4 $cm^{-1}$, showing a multilayer $PtS_2$. The inset of Figure 1(d) shows the Raman signature of crystalline Si at 520 $cm^{-1}$. Figure 1(e) demonstrates the AFM image of a $PtS_2$ nanoflake on the Si substrate, and Figure 1(f) shows a corresponding step profile height that was obtained from the AFM line scan as indicated. The thickness of $PtS_2$ nanoflake is up to ∼40 nm. The mix-dimensional (2D/3D) $PtS_2$/pyramids Si p-n junction was established by exfoliating the multi-layer $PtS_2$ nanoflakes on pyramids silicon substrates. We explored the electrical properties of $PtS_2$/pyramids Si heterostructures at 300 K as well as at low temperatures up to 50 K. Here, we examined the annealing effect on $PtS_2$/pyramids Si junction. First of all, we studied the ($I_{DS}$-$V_{DS}$) characteristics without annealing the interface between $PtS_2$ and pyramids Si, and the obvious device seems to act in a diode-like behavior, however, when the same device was annealed for 4 hours at 200 °C in the furnace tube under the flow of Ar/$H_2$ (97.5% Ar & 2.5% $H_2$), electrical characteristics were enhanced due to the clean interface of $PtS_2$/pyramids Si p-n junction. However vacuum annealing showed better performance.

The annealing effect on p-n junction performance is shown in Figure S2 of supporting information. Further, post-annealing measurements were performed on all devices. The J-V features of intermolecular forces of heterojunction are demonstrated by the following relation: [25-27]

$$J \propto exp^{\frac{qv}{nkT}}$$

where T indicates the temperature, n represents the ideality factor, q is the charge, and k is the Boltzmann constant. The recombination current density occurs as a result of the charge carriers that are present in the space charge region close to the boundary in the majority of the p-n junctions, whereas the diffusion current density, due to diffusion of charge carriers in the neutral regions, lead to forward dark current density.[26, 28, 29] Figure 2(a) shows the energy band diagram before contact. When n-type PtS$_2$ flake is transferred on a p-type substrate by using the transparent Polydimethylsiloxane (PDMS) stamp method, electrons move from n-PtS$_2$ into p-Si at the interface due to the higher Fermi level of PtS$_2$. A p-n junction is made at the interface as a result of the alignment of Fermi levels which stop the further flow of electrons. At zero bias, both the holes and the electrons would not go across the junction due to alignment of the Fermi levels. Figure 2(b) shows the band diagram of the PtS$_2$/pyramids Si heterojunction diode after contact. Both electrons and holes can move across the junction at forward bias. At reverse bias, electrons and holes would not move across the junction due to an increase in barrier height. Devices show rectifying behavior under forward bias and reverse bias at 300 K in Figure 2(c). The ideality factor can be extracted in the forward-biased regime of output ($I_{ds} - V_{ds}$) characteristics by using the diode equation as follows: [27, 30]

$$\ln(I_D) = \ln(I_S) + \left(\frac{q}{nK_B T}\right) V$$

where $I_D$ (diode current in forwarding bias regime) and $I_S$ (diode reverse-bias saturation current), $V$ (applied voltage across the diode), $n$ (ideality factor), $q$ (charge on the electron), $T$ denotes the temperature (300 K), and $K_B$ (Boltzmann's constant). The slope of the transfer plot in forwarding bias gives $q/{nK_B T}$, from which $n$ is calculated. The value of the ideality factor of the diode was found to be 1.5 as shown in Figure 2(d).

To check the breakdown mechanism in PtS$_2$/pyramids Si junction, ($I_{DS}$-$V_{DS}$) characteristics at different temperatures were measured for several devices. Devices demonstrated rectifying behavior under forward and reverse biases at all temperatures. The breakdown region of the PtS$_2$/pyramids Si diode can be seen in the negative region of Figure 3(a). It exhibited different breakdown phenomena, such as Zener tunneling diode under reverse bias. Generally, both Zener tunneling, and avalanche breakdown diodes refer to the breakdown of a junction under the action of reverse biasing. Particularly, if the diode is made to operate in the reverse bias, a minor current passes across it, and the diode is considered as an open circuit.[31] The reverse bias current increases abruptly if we keep on increasing the reverse bias voltage to the point where the dynamic resistance becomes low. We can also designation this reverse bias voltage as the breakdown voltage, and it is independent of the reverse current following through a p-n diode.[31] Consequently, under reverse bias, both breakdowns, either Zener or avalanche, may exist. The breakdown criteria for Zener tunneling diodes, the breakdown voltage should be less than 4E$_g$/e, for band-to-band tunneling, where E$_g$ is the bandgap and e is the charge of the electron.[28, 32] The bandgap of Si is 1.1 eV,[26] while for PtS$_2$ its value varies from 0.3 eV & for E$_g$/e for Si and PtS$_2$ the value is 4.4 and 1.2 V, respectively. The band-to-band tunneling mechanism is predominant at lower voltage with heavy doping in a thin depletion region, where quantum tunneling occurs, which causes current to flow while avalanche breakdown dominates at a higher voltage with a low doping level resulting in a longer depletion region.[31] To ensure the mechanism of band-to-band tunneling and avalanche breakdown, ($I_{DS}$-$V_{DS}$) characteristics were observed at a various temperatures from 50 to 300 K. For all the devices, the junction shows Zener tunneling at all temperatures in the range of small breakdown voltage between -1.0 and -6 V (in Figures 3(a)), while avalanche breakdown only at higher temperatures from 200 K to 300 K with high breakdown voltage from -15 to –12 V (in Figures 3(b)). We found that breakdown voltage (-6 to -1.0 V) is the negative temperature coefficient for the Zener diode, indicating that it decreases with the temperature increases, as shown in Figures 3(a). So, we can also name the breakdown as Zener breakdown.[28, 31, 32] On the other hand, we found that for avalanche breakdown (-15 to -12 V), breakdown voltage increases as the temperature increases, suggesting that breakdown voltage is the positive temperature coefficient as shown in Figures 3(b). Consequently, such a breakdown can be named as avalanche breakdown.[28, 31, 32] Also, in the case of avalanche breakdown, the devices have no stable reverse voltage at any temperature, as shown in Figures 3(b), because avalanche breakdown is caused by the impact of ionization, which may have a low probability at a low temperature. As in the case of avalanche breakdown, the reverse breakdown voltage is a positive temperature coefficient, so we may observe a high breakdown voltage by increasing the range of reverse bias voltage. But too high a voltage may produce an unrecoverable breakdown in a p-n junction. Oppositely, the p-n diode shows a more stable current for Zener breakdown when compared to avalanche breakdown for all showed temperature ranges, as presented in Figure 3(a). Since the Zener effect results from the tunneling mechanism and the sample does not allow the charge carriers to pass after a breakdown occurs. Despite the fact avalanche breakdown is due to the multiplication of ionization due to the collision of charge carriers, and the material continues to conduct even if reverse bias voltage drops below the breakdown voltage.[33, 34] For photodetectors, a balanced photocurrent is essential, so utilizing Zener tunneling is more suitable, [35] although for tunnel diodes avalanche break down is more appropriate. Hence, the working of a p-n junction in Zener breakdown regimes can be used as a photodetector due to its function as a voltage regulator. The relation between Zener and avalanche breakdown voltages as a function of temperature is shown in Figure 3(c). The energy band diagrams for a p-n diode at zeros bias, at Zener tunneling, and avalanche breakdown are shown in Figures 3(a, b, c), respectively.

Moreover, we further investigated why the $PtS_2$/pyramids Si junction showed amazing photo conducting performance under the light of different intensities ($\lambda$= 500 nm). The separation rate of electron-hole pair increases as a result of an increase in the external field till all charge carriers contribute to the photocurrent before the recombination occurs, and subsequently, saturation results as shown in the reverse biased region of Figure 4(a). The maximum values for both the open-circuit voltage ($V_{OC}$) and the short-circuit current ($I_{SC}$) were observed to be 0.45 V, and 10 $\mu$A, respectively, at input intensity of light 70.32 mW/cm$^2$ (in Figure 4(b)). Figure 4(c) shows the output power of the p-n diode as a function of $V_{OC}$. It can be seen that $PtS_2$/pyramids Si p-n diode displayed maximum output power for an input light with greater intensity. Moreover, to evaluate the p-n junction photoresponse behavior, the time-dependent current under light with constant $V_{ds}$= 1V was also measured for $PtS_2$/ Si p-n diode with and without pyramids, as shown in Figures 4(d), where the device displayed the current switching behavior when the light was switched on and off and greater photoresponse was detected for a p-n diode with pyramids. We confirmed the photoresponse performance by calculating the photocurrent, Photogain (G), and photoresponsivity ($R_{ph}$) by using the following equations $I_{ph} = I_{photo}$ - $I_{dark}$, $G = R_{ph}\frac{hc}{q\lambda}$ and $R_{ph} = I_{ph}$ /PA, [36-38] respectively, where $I_{photo}$ is the current under light illumination and $I_{dark}$ is current in dark, P is the intensity of light, and A indicates effective area. In our p-n junction, we find high a value of responsivity 11.88 A/W with a photo gain of 67.10 as compared to the previous reported p-n junction based on growth with Si.[35]

## 4.0 Conclusion and Summary:

In summary, to fabricate the $PtS_2$/pyramids Si junctions, we exfoliated multi-layer $PtS_2$ nanoflakes on a p-type silicon via mechanical exfoliation. The p-n junction exhibited rectifying features over a wide voltage. We have also demonstrated that p-n junctions show improved electrical characteristics with annealing. At various temperatures, it is demonstrated that in the low reverse bias, the Zener breakdown tunneling occurs with a negative temperature coefficient to breakdown voltage. On the other hand, for a regime with greater breakdown voltage showing a positive coefficient of temperature. So, this breakdown may be ascribed to the avalanche breakdown diode. The p-n junctions exhibit photovoltaic characteristics, with high photoresponsivity and photo gain of $R_{ph} = 11.88 A/W$, $G = 67.10$, respectively. Such a high-performance p-n junction created a strong electric field within small reverse-biased voltage, and thus it showed an effective separation of electron-hole pairs of photo-generated charge carriers. This study recommends innovative and promising applications of $PtS_2$/ pyramids Si in different electronics, such as light-emitting diodes, solar cells, Zener diodes, photodetector, and tunnel diodes.

## 5.0 Acknowledgments:

This work made use of the KECK II, EPIC facility of Northwestern University's NUANCE Center, which has received support from the SHyNE Resource [NSF Grant ECCS-2025633], Northwestern's MRSEC program (NSF Grant DMR-1720139), the Keck Foundation, and the State of Illinois through IIN. MZI would like to acknowledge Higher Education Commission (HEC) under the National Research Program for Universities (NRPU) with project no. HEC/R&D/NRPU/2017/7876. MWI acknowledge funding project number R-ORIC-21/FEAS-10.

**Note:** The data that support the findings of this study are available within the article.

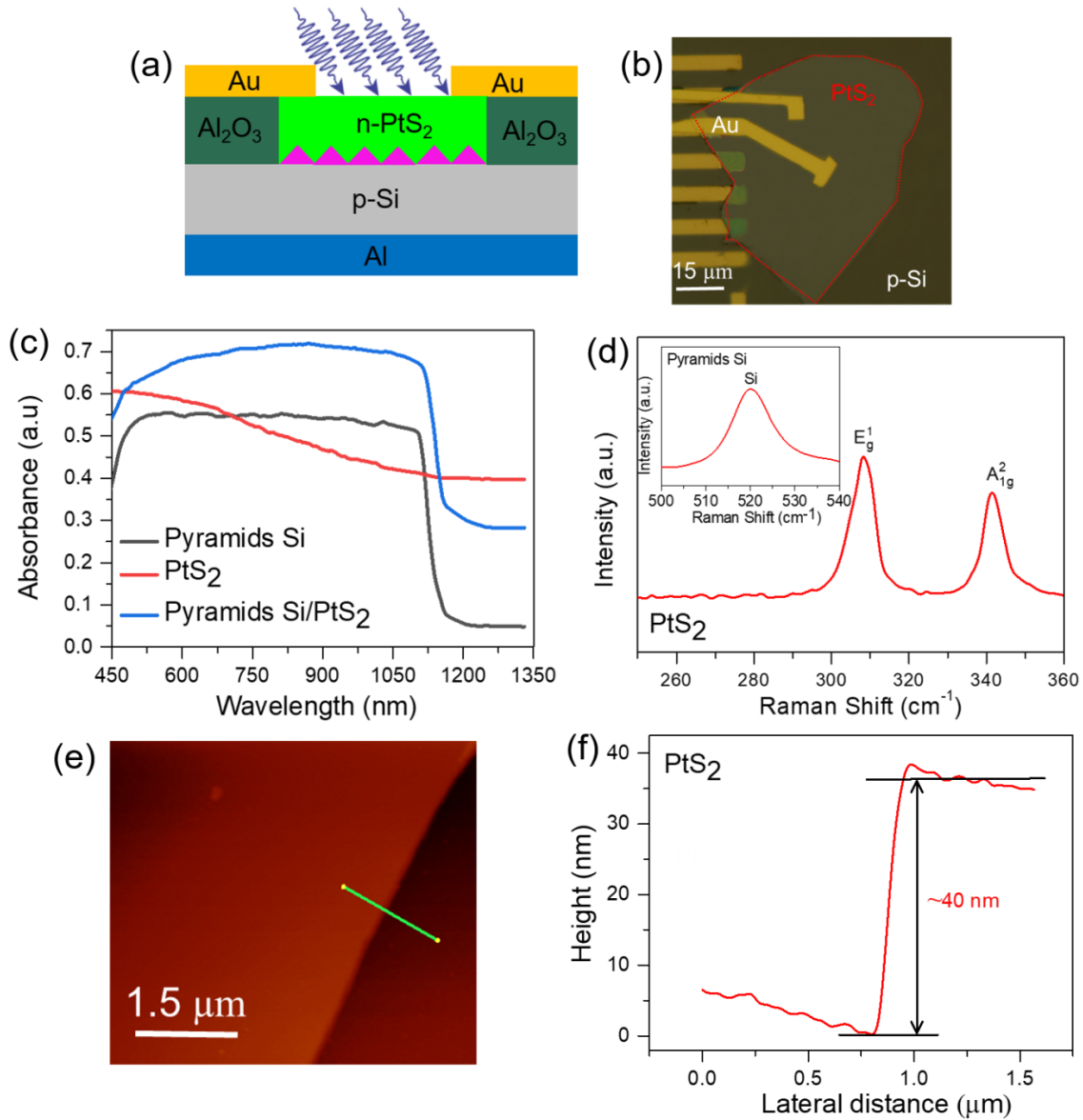

**Figure 1.** (a) Diagram illustration of PtS$_2$/pyramids Si van der Waals heterojunction diode. The top Au and bottom Al contacts were used for electrical transport measurements. (b) Final optical image of the device. The p-type substrate acts as a source while PtS$_2$ acts as a drain with a 20 nm Al$_2$O$_3$ layer. (c) Absorption spectra of the PtS$_2$ pyramids Si and PtS$_2$/pyramids Si heterostructure. (e) Raman spectrum of the multi-layered PtS$_2$ nanoflake on a p-type Si substrate. The inset of c shows the Raman spectrum for Si. (e) AFM image of PtS$_2$ nanoflake. (f) The thickness of PtS$_2$ nanoflake is 40 nm from AFM as scanned.

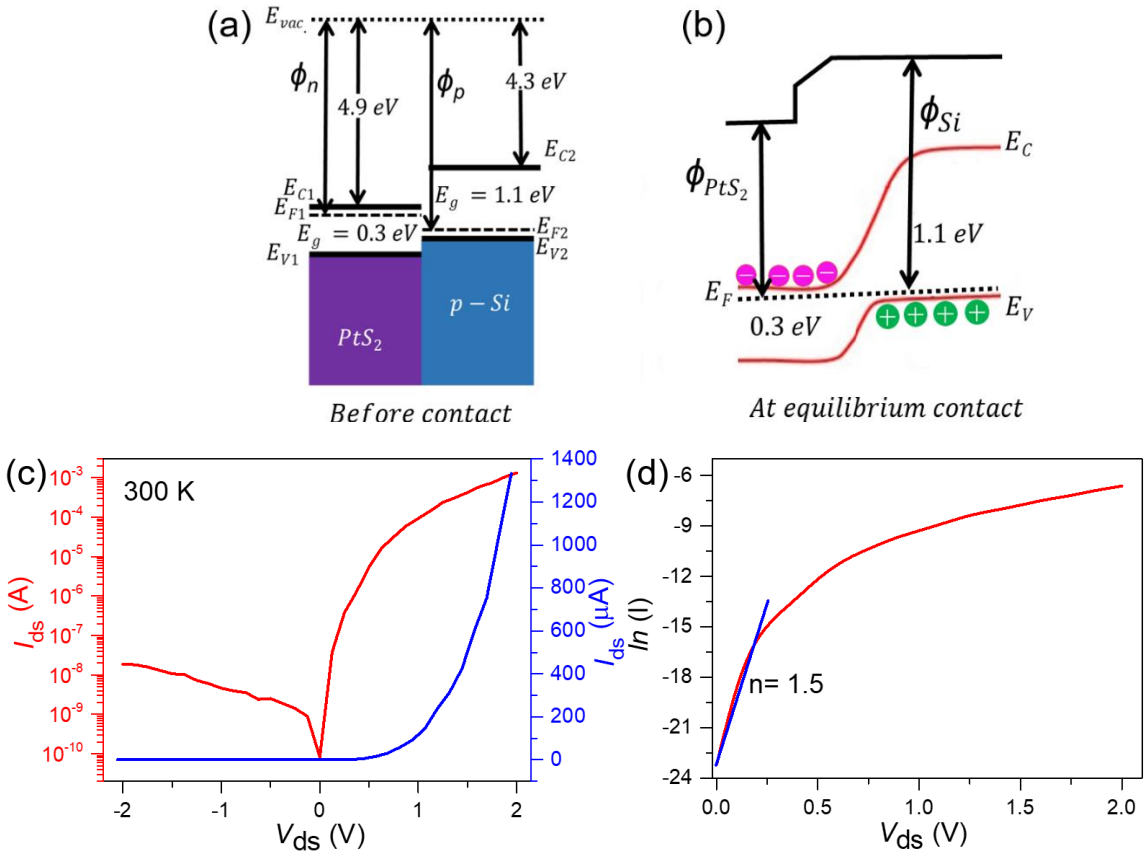

**Figure 2.** (a) Schematic of an energy band diagram of PtS$_2$ and Si before contact. (b) Schematic of the energy band diagram of PtS$_2$ and Si at equilibrium contact. (c) Rectifying effect of PtS$_2$/pyramids Si heterojunction diode at zero gate bias. (d) ($I_{DS}$-$V_{DS}$) characteristic in the forward-biased region shows the value of the ideality factor.

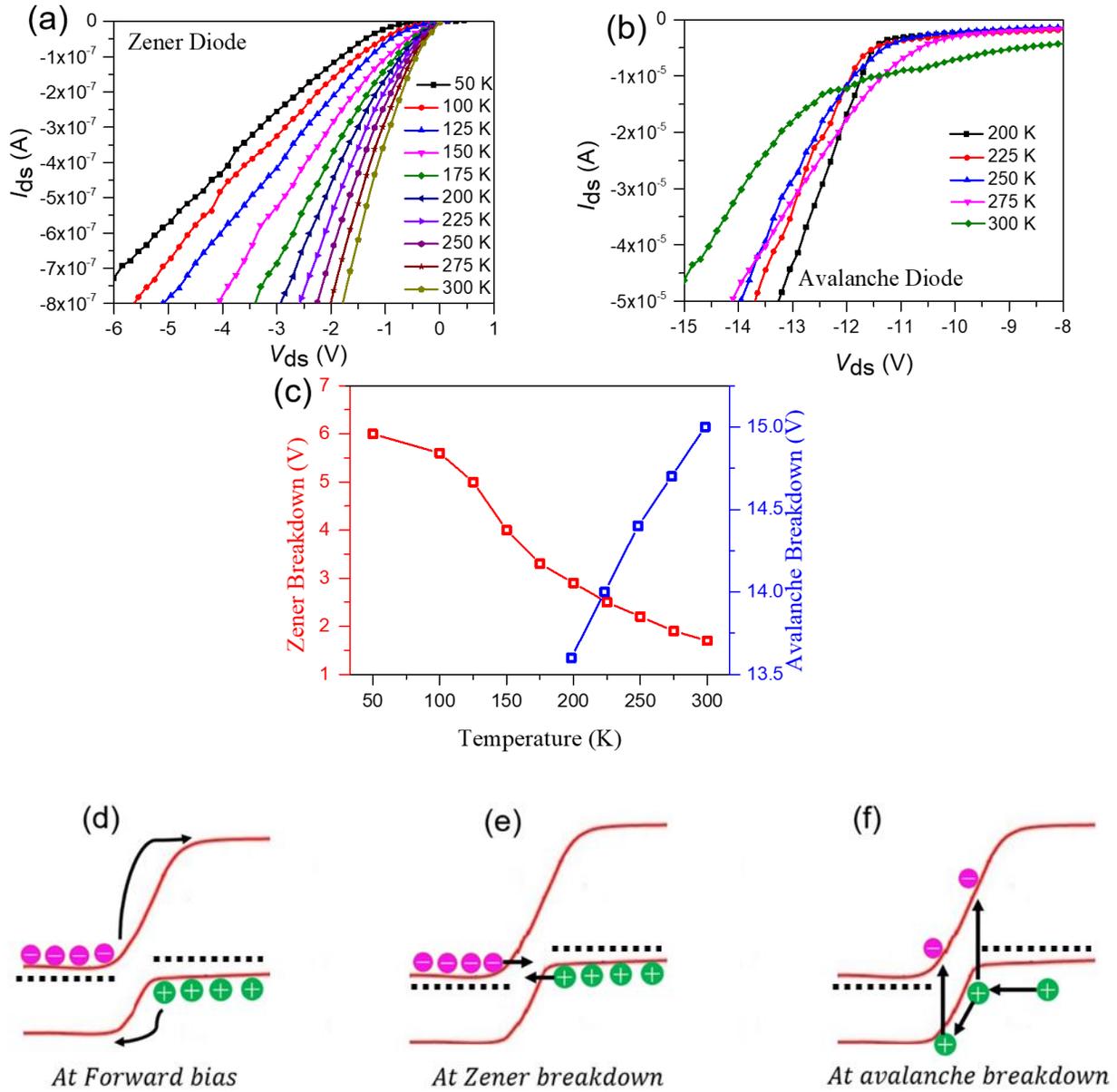

**Figure 3.** (a) Extend ($I_{DS}$-$V_{DS}$) characteristics viewing Zener diode. (b) Extend ($I_{DS}$-$V_{DS}$) characteristics viewing avalanche diode. (c) Dependence of Zener and avalanche breakdown voltages on temperature. Schematic of the energy band diagram of PtS$_2$ and Si at (d) forward bias, (e) Zener breakdown, and (f) avalanche breakdown.

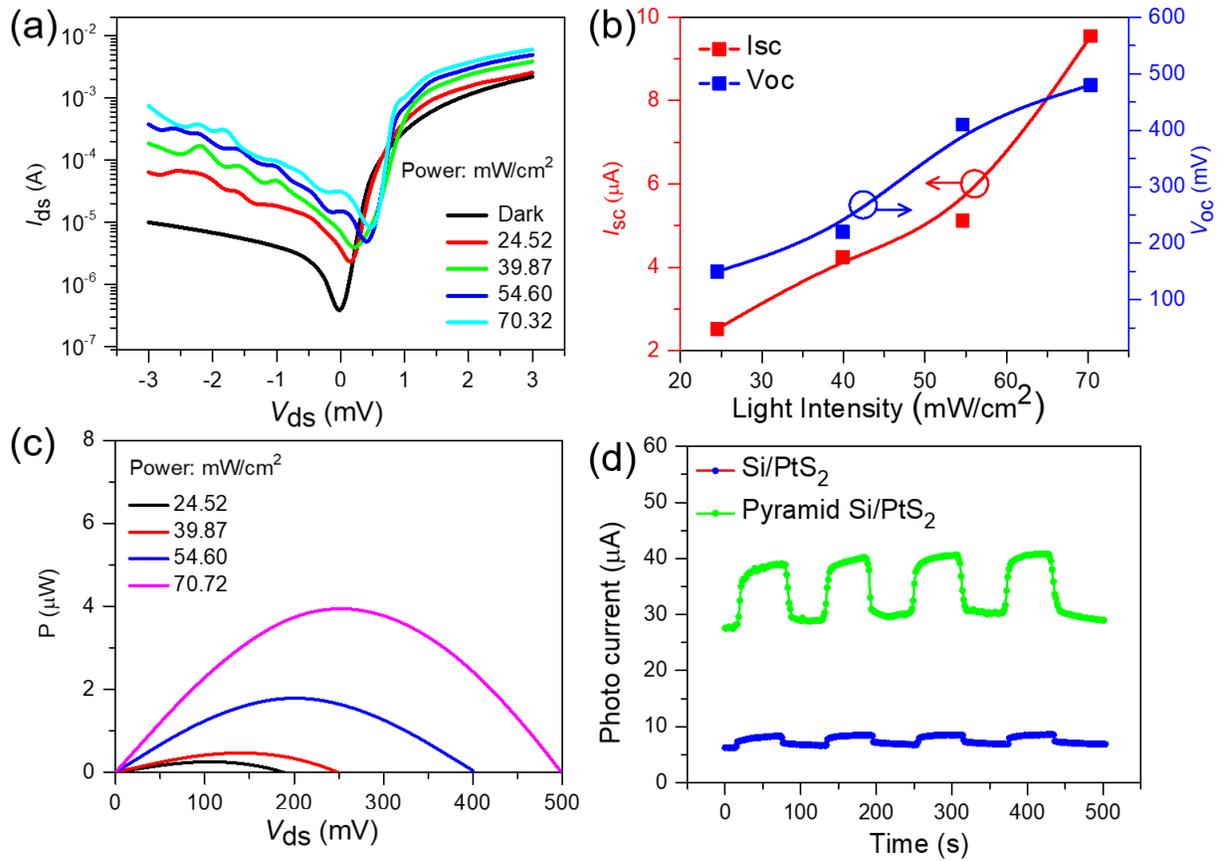

**Figure 4.** (a) ($I_{DS}$-$V_{DS}$) characteristics of PtS$_2$/pyramids Si heterojunction diode under the dark and on illuminating with the light of different intensities ($\lambda$= 500 nm). (b) $V_{oc}$ and $I_{sc}$ as a function of input power. (c) The electrical power of PtS$_2$/pyramids Si heterojunction diode with different input power. (d) Photoresponse of PtS$_2$/pyramids Si heterojunction diode under illuminations PtS$_2$/pyramids Si and PtS$_2$/ Si heterojunctions.

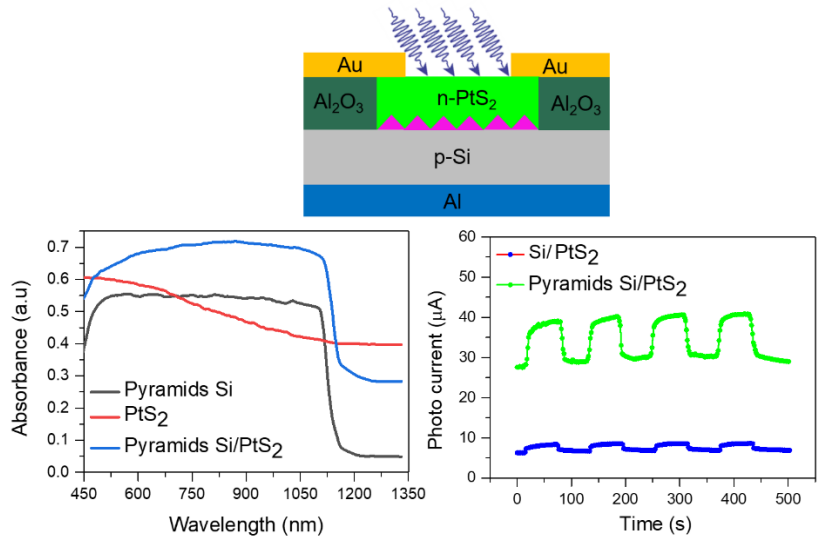